\documentclass[aps,prd,twocolumn,eqsecnum,amssymb,amsmath,showpacs,a4paper, superscriptaddress]{revtex4-2}

\usepackage{graphicx}
\usepackage{amsfonts}
\usepackage{amsmath}
\usepackage{hyperref}
\usepackage{units}
\usepackage{color}
\usepackage{dcolumn}
\usepackage{bm}
\usepackage{float}
\usepackage{cleveref}
\usepackage[normalem]{ulem}
\usepackage{appendix}
\usepackage{mathtools}
\usepackage{esvect}

\usepackage[colorinlistoftodos, size=tiny, bordercolor=white]{todonotes}
\usepackage{comment}

\graphicspath{ {images/} }
\usepackage{mathrsfs}
\usepackage{amssymb}
\usepackage{dsfont}
\usepackage{enumitem}
\usepackage{gensymb}
\usepackage{bm}

\usepackage{mathtools}
\usepackage{chngcntr}
\counterwithout{equation}{section}

\makeatletter
\let\cat@comma@active\@empty
\makeatother

\newcommand{\h}{\mathrm{H}}

\usepackage{xr}

\begin{document}

\title{On the origin of quasinormal modes in semi-open systems}

\author{Leonardo Solidoro}
\email{leonardo.solidoro@nottingham.ac.uk}
\affiliation{School of Mathematical Sciences, University of Nottingham, University Park, Nottingham, NG7 2RD, UK}
\affiliation{Nottingham Centre of Gravity, University of Nottingham,
University Park, Nottingham NG7 2RD, UK}

\author{Sam~Patrick}
\affiliation{Department of Physics, King’s College London, University of London, Strand, London, WC2R 2LS, UK}

\author{Silke Weinfurtner}
\affiliation{School of Mathematical Sciences, University of Nottingham, University Park, Nottingham, NG7 2RD, UK}
\affiliation{Nottingham Centre of Gravity, University of Nottingham,
University Park, Nottingham NG7 2RD, UK}
\affiliation{Centre for the Mathematics and Theoretical Physics of Quantum Non-Equilibrium Systems, University of Nottingham, Nottingham, NG7 2RD, UK}

\author{Ruth Gregory}
\affiliation{Department of Physics, King’s College London, University of London, Strand, London, WC2R 2LS, UK}
\affiliation{Perimeter Institute, 31 Caroline Street North, Waterloo, ON, N2L 2Y5, Canada}

\date{\today}
\begin{abstract}
\noindent
Astrophysical black holes are open systems  which, when perturbed, radiate quasi-normal modes (QNMs) to infinity. By contrast, laboratory analogues are necessarily finite-sized, presenting a potential obstacle to exciting QNMs in experiments. We explore how the QNM spectrum of a toy-model black hole changes when enclosed by a partially reflecting wall with adjustable reflectivity. Our results reveal a continuous connection between the QNM spectra of open and finite-sized systems. Additionally, we demonstrate that QNMs in this setup are easily excited by incoherent background noise. This work opens new avenues for studying QNMs of black holes and compact objects in laboratory settings, where finite-size effects and noise are unavoidable.
\end{abstract}
\maketitle

\emph{Introduction.---}The success of the black hole spectroscopy program is founded on the use of gravitational waves produced by merging binaries as a probe of the system's properties \cite{abbott2016}. At late times during the so-called \emph{ringdown} phase, the remnant emits a characteristic signal as it sheds energy and relaxes into its final state. This signal consists of a discrete set of damped harmonic oscillators referred to as quasinormal modes (QNMs) \cite{Ferrari1984,Echeverria1989,kokkotas1999quasi,Cardoso2009,Berti2009, Konoplya:2011qq}, which are natural resonances of linear open systems. 
Next generation gravitational wave detectors such as LISA and Einstein telescope will profit from enhanced sensitivity \cite{amaro2017,Maggiore2019}, making it paramount to understand how the ringdown is influenced by various high-order effects, e.g.\
environmental factors \cite{cannizzaro2024impact} and nonlinearities \cite{Cheung2023}.

Recent investigation into the spectral stability of black holes has demonstrated the susceptibility of the QNM spectrum to small, local perturbations to the scattering potential experienced by gravitational waves around a black hole
\cite{Nollert1996,Aguirregabiria1996,nollert1999quantifying,Cheung2022}. In some cases, the spectrum can even admit additional long-lived modes called quasi-bound states, potentially leading to higher-order effects as a result of their long lifetimes \cite{witek2013,Yang2021}.
Understanding such modifications of the QNM spectrum may have important consequences for the black hole spectroscopy programme \cite{jaramillo2021pseudospectrum,jaramillo2022gravitational}.
In this regard, it is natural to consider non-perturbative scenarios where energy is retained by the system due to an effective reflection mechanism acting on escaping radiation. Such situations arise naturally in gravity in the context of neutron stars \cite{kokkotas1999quasi}, exotic compact objects \cite{pani1,pani2,pani3,pani4} and asymptotically anti-de Sitter spacetimes \cite{Berti2009, Konoplya:2011qq}.

Gravity simulators present an ideal platform where such an enquiry can be pursued. These are physical systems in which excitations behave as though they propagate on a curved spacetime, permitting the experimental investigation of elusive phenomena like Hawking radiation \cite{weinfurtner2011,euve2016,munoz2019observation} and rotational superradiance \cite{visser2005,Torres2017,solnyshkov2019,braidotti2022,cromb2020}. Notably, ringdown signals have been studied in polariton superfluids \cite{jacquet2023}, optical solitons \cite{Burgess2024} and in hydrodynamic systems \cite{Patrick2018,Torres2020}.
Although treating these platforms as effectively open systems is usually justified by a combination of energy dissipation, limited temporal evolution and engineering of absorptive boundaries, a complete analysis should not neglect finite-size effects resulting from experimental confinement mechanisms, e.g. in classical fluids \cite{Torres2020}, superfluids \cite{svancara2024}, cold-atomic-gases \cite{kolobov2021observation}, polariton fluids \cite{delhom2024entanglement} etc.

All of this serves to emphasise the importance of understanding how finite-size effects modify the QNM spectrum.
In this Letter, we study the effect of a partially reflective boundary on the QNM spectrum of the P\"oschl-Teller (PT) potential.
This system is sufficiently versatile to capture qualitative features of potentials around astrophysical black holes and their analogues \cite{Berti2009, Konoplya:2011qq,Ferrari1984,Torres2022,Burgess2024} whilst remaining analytically solvable. We demonstrate that finite-size effects endow the system with an additional control parameter that can facilitate the experimental detection of QNMs, enabling control of their lifetimes and their excitation by experimental noise.

\emph{General set-up.---}A simple system which captures the essential physics of QNMs is the one-dimensional wave equation with a repulsive potential $V(x)$ \cite{kokkotas1999quasi,Berti2009, Konoplya:2011qq},
\begin{align}
\label{eq:WaveEq}
\left(\frac{\partial^2}{\partial t^2} - \frac{\partial^2}{\partial x^2} + V(x)\right)\Psi = 0\,,
\end{align}
where the wave speed is set to unity and we assume \mbox{$V\to 0$} as $x\to\pm\infty$.
QNMs are resonances of open systems, i.e. systems that lose energy, in this case, to spatial infinity. By looking for solutions of the form $\Psi(x,t) = e^{-i\omega t}\psi(x)$, the wave equation reduces to a Schr\"odinger-like equation $\psi^{\prime\prime} + \left[\omega^2-V(x)\right]\psi = 0\,,$
and the set of QNMs are found as the set of frequencies whose related modes satisfy the purely out-going boundary conditions,
\begin{align}
\label{eq:FreeBound}
    \lim_{x\to \pm\infty}\psi \propto e^{\pm i\omega x}. 
\end{align}
More generally, the problem of calculating QNMs can be formulated as a scattering problem. Consider a solution of \eqref{eq:WaveEq} consisting of an incident and a reflected contribution on the right and a transmitted one on the left,
\begin{align}
\label{eq:scatt}
    \psi(x) =
    \begin{cases}
        A_i\,e^{-i\omega x} + A_r\,e^{+i\omega x}\quad &,\,\,x\to +\infty\,, \\
        A_t\,e^{-i\omega x}\quad &,\,\,x\to -\infty
    \end{cases}\,.
\end{align}
From this, one can define the reflection and transmission coefficients $R(\omega) = A_r/A_i$ and $T(\omega) = A_t/A_i$ respectively,
and one sees that the QNM frequencies determined by \eqref{eq:FreeBound} are poles of the reflection coefficient \cite{Berti2009}. 

Consider now the semi-open problem where a wall of reflectivity $\varepsilon$ is situated at $x_b$ such that $V(x_b)$ is effectively zero.
The boundary condition becomes,
\begin{align} \label{eq:bc2}
    \psi(x\to x_b)\propto  \varepsilon e^{i\omega (x-x_b)} + e^{-i\omega (x-x_b)} \,.
\end{align}
We choose a constant $\varepsilon \in[0,1]$, allowing us to vary the amount of absorption at the wall: for $\varepsilon = 1$ the wall is fully reflective whilst for $\varepsilon = 0$ we recover the open problem \footnote{Note that our choice of $\varepsilon$ real and positive corresponds to a Neumann boundary condition in the case of a complete reflection, whereas choosing $\varepsilon$ negative/complex would lead to a Dirichlet/Robin boundary condition. In general, $\varepsilon$ could depend on $\omega$ but for the sake of simplicity, we model it here as a constant.}.
The compatibility of \eqref{eq:bc2} with \eqref{eq:scatt} implies the following resonance condition \cite{macedo2018spectral},
\begin{align}
\label{eq:ResCond}
    \varepsilon R(\omega) e^{2i\omega x_b} = 1\,.
\end{align}
The QNM condition is retrieved in the $\varepsilon\to 0$ limit, whilst for a finite reflectivity one should expect the resonances to deviate from the ones of the open system. This is well understood in the context of spectral stability analysis, where a small perturbation of $V$ can significantly modify the spectrum \cite{jaramillo2021pseudospectrum}.
The continuous evolution of the resonances in the complex plane as $\varepsilon$ increases can be described in terms of integral curves of an ordinary differential equation \cite{dolan2024}. The resonance condition \eqref{eq:ResCond} can be expressed as $\varepsilon - R^{-1}(\omega_n^\varepsilon) = 0$, where $\omega_n^\varepsilon$ are the resonant frequencies indexed by $n=0,1,...$, and the superscript indicates the value at a particular $\varepsilon$. By differentiating with respect to the reflectivity, one gets,
\begin{align}
\label{eq:trajectory}      
    \frac{\partial\omega_n}{\partial\varepsilon} = -\frac{R^2}{R^{\prime}}\bigg|_{\omega=\omega_n} \equiv \h(\omega_n)\,,
\end{align}
where $R^\prime = \partial_\omega R$. The right-hand side defines a vector field along which the real and imaginary parts of the resonances evolve as we change the boundary's reflectivity.

\emph{P\"oschl-Teller potential.---}We now apply this framework to the P\"oschl-Teller (PT) potential,
\begin{align}
\label{eq:PTpot}
V_\mathrm{pt}(x) = \frac{V_0}{\cosh^2(\alpha\, x)}\,.
\end{align}
In this case, \eqref{eq:WaveEq} has an analytical solutions in the form $e^{-i\omega t}\psi(x)$.
Defining a new variable $y = 1/(1+e^{-2\alpha x})$, the spatial part can be written as,
\begin{align}
    \label{eq:ptsolution}
   \psi(y) &=  (1-y)^{-i\omega/2\alpha}\left[A\,y^{-i\omega/2\alpha}{}_2F_1(a,b,c;y)+\right.\nonumber   \\
   &\left.+B\,y^{i\omega/2\alpha}{}_2F_1(1+a-c,1+b-c,2-c;y)\right]\,,
\end{align}
where $A,B$ are constants and ${}_2F_1$ hypergeometric functions \cite{Florentin1966} whose parameters are given by,
\begin{equation}
    a,b =\frac{1}{2} -i\omega/\alpha
     \pm \frac{1}{2}\sqrt{1-\frac{4V_0}{\alpha^2}}, \qquad 
    c = 1-i\omega/\alpha\, ,
\end{equation}
with the $+$ ($-$) sign for $a$ ($b$). Without loss of generality, from now on we take $V_0 = 1$, so that the problem is characterised by the two dimensionless scales $\alpha$ and $x_b$.
To impose the resonance condition \eqref{eq:ResCond}, we need to define the reflection coefficient.
Assuming that $V(x_b)$ is close to vanishing, \eqref{eq:ptsolution} is well-approximated by its asymptotic expansion and the in- and out-going wave contributions can be identified (see Appendix A).
Their ratio gives the reflection coefficient, 
\begin{align}
    \label{eq:ptref}
    R
     =  \frac{\Gamma(a)\Gamma(b)}{\Gamma(c-a)\Gamma(c-b)} 
    \frac{\Gamma(i\omega/\alpha)}{\Gamma(-i\omega/\alpha)}.
\end{align}
Poles of this expression give the QNMs of the open system,
\begin{align}
\label{eq:PTqnms}
\omega^{\varepsilon=0}_n &= \pm \sqrt{1-\alpha^2/4} - i\alpha (2n+1)/2 \,,
\end{align}
where $n$ is called the overtone number. All QNM frequencies of the PT potential share the same real part and differ only in their imaginary part.
The mode with $n=0$ is called the fundamental mode, since it has the longest lifetime, whereas overtones with $n>0$ are characterised by stronger damping \cite{Berti2009}. This particular feature, also relevant for astrophysical black holes, renders the observation of overtones difficult. 
Consequently, the analysis of the systems' properties is often limited to the detection of the fundamental mode.

\begin{figure*}
    \centering
    \includegraphics[width=0.95\linewidth]{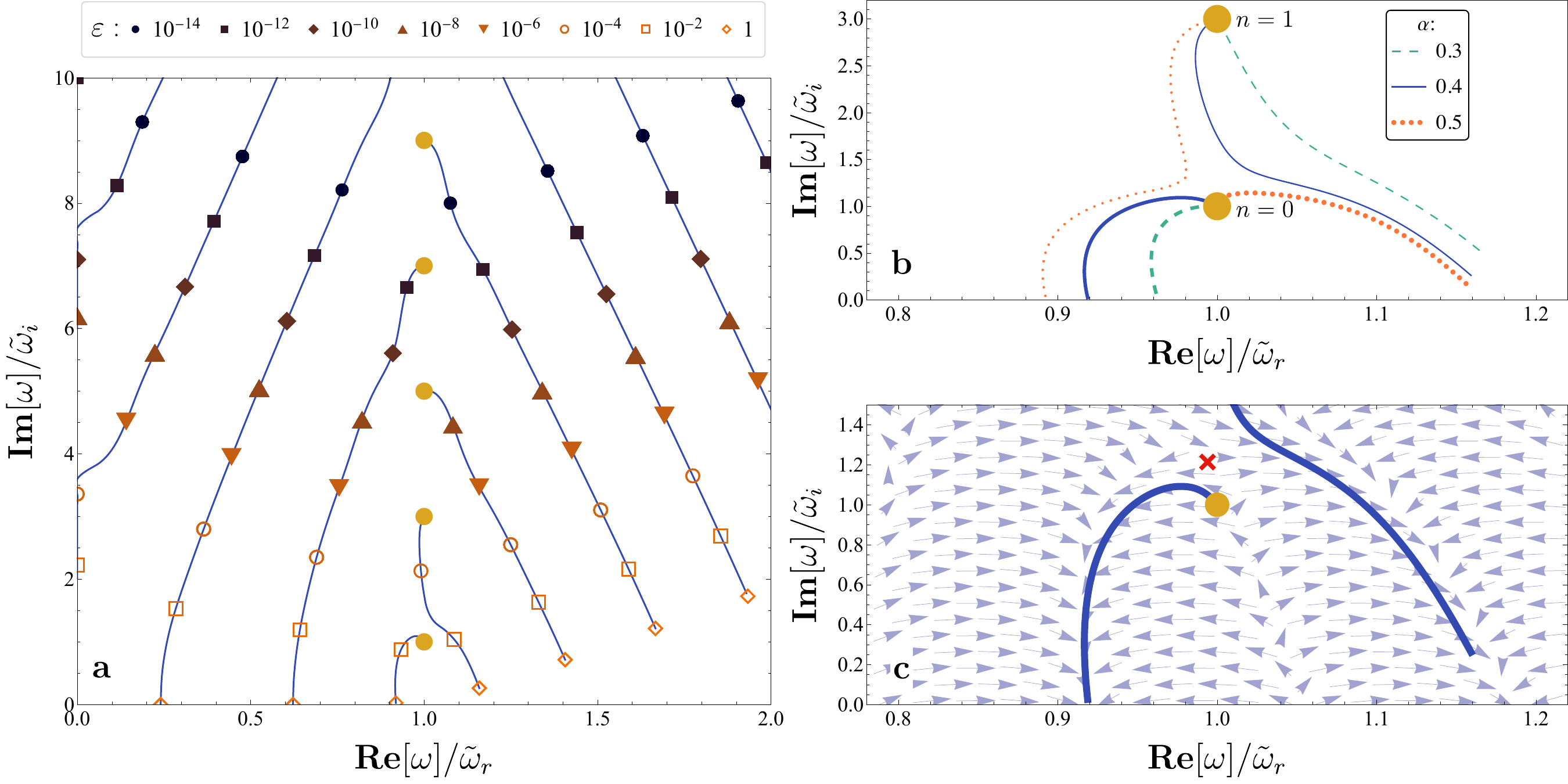}
    \caption{\textbf{a.} Trajectories in the complex plane of the first 11 resonances of the PT potential with a partially-reflective barrier relative to the fundamental QNM $\omega_{n=0}^{\varepsilon=0}\equiv \tilde{\omega}_r +i \tilde{\omega}_i$. Here we take $\alpha = 0.4$ and the position of the wall is fixed at $x_b = 10$. The large yellow dots represent QNMs of the open system, whereas the blue lines are obtained varying $\varepsilon$. Coloured markers label the resonances at particular values of reflectivity. \textbf{b.} Trajectories of the resonances as reflectivity increases, for $\alpha = 0.3$ (Green, dashed), $=0.4$ (Blue, solid), $=0.5$ (Orange, dotted) and $x_b = 10$. The yellow dots are $\omega_{n=0,1}^{\varepsilon=0}$ and the lines end at $\varepsilon = 1$. \textbf{c.} The migration of the resonance is compared against the vector field $\h(\omega)$ for $\alpha=0.4$. The red cross is placed on the repelling point, i.e. the pole of $\h$.}
    \label{fig:QNMdrift}
\end{figure*}

Resonances of the semi-open system are obtained by numerically solving \eqref{eq:ResCond}.
In Fig.~\ref{fig:QNMdrift}a, we display their continuous migration in the complex plane as $\varepsilon$ is varied, fixing $x_b = 10$ and $\alpha = 0.4$. The trajectories are presented relative to the fundamental QNM $\omega_{n=0}^{\varepsilon=0}\equiv \tilde{\omega}_r +i \tilde{\omega}_i$. Large yellow dots represent QNMs of the open system for $\varepsilon=0$ and, as $\varepsilon$ is increased, the resonances migrate down toward the real line.
We can identify three qualitatively different types of QNM when $\varepsilon=1$.
Modes with $\mathrm{Re}[\omega]=0$ correspond to purely damped solutions (discussed in \cite{dolan2024}).
Modes with $0<\mathrm{Re}[\omega]<\tilde{\omega}_r$ are damped oscillatory modes which are localised between the hard wall and the barrier $V$. These are quasibound states (QBS) which need to tunnel through $V$ to escape to $x\to-\infty$, leading to long lifetimes.
Finally, modes with $\mathrm{Re}[\omega]>\tilde{\omega}_r$ pass over the barrier and experience only a small amount of reflection due to the inhomogeneous nature of $V$.
Consequently, they are more strongly damped and can be identified with the new overtones of the system.
These overtones possess distinctly different real frequencies and an imaginary part which is reduced significantly relative to the open system.
Note, we are mainly interested in the regime of small $\alpha$ (wide potentials) where the decay rate of $\omega^{\varepsilon=0}_{n=0}$ is an order of magnitude smaller than its oscillation frequency, which is the astrophysically relevant case \cite{leaver1985analytic}. 
The scale $x_b$ then determines the number of QBS in semi-open system.

In Fig.~\ref{fig:QNMdrift}b, we provide a closeup of the $n=0,1$ modes and show the effect of changing $\alpha$, i.e. the width of the PT potential.
As $\alpha$ gets larger, the potential barrier becomes narrower and more QBSs can fit in the cavity formed by the hard wall and potential barrier.
In the complex $\omega$ plane, one would then see all the $\omega^{\varepsilon=1}_n$ moving toward lower $\mathrm{Re}[\omega]$ and the trajectories emerging from $\omega^{\varepsilon=0}_n$ orbiting around the yellow points in Fig.~\ref{fig:QNMdrift}a in the clockwise direction.
However, at critical values of $\alpha$, there can be discontinuous jumps where the migrations lines ending at $\omega^{\varepsilon=1}_n$ change which of the $\omega^{\varepsilon=0}_n$ they are connected to.
These discontinuous jumps are related to repelling points of \eqref{eq:ResCond}, illustrated as the red cross in Fig.~\ref{fig:QNMdrift}c.
A migration line is an integral curve of the vector field $\h(\omega)$, defined in \eqref{eq:trajectory}, and is deflected either to the left or to the right of the repelling point, i.e. the poles of $\h(\omega)$, as the resonances move down toward the real axis.
A small change in $\alpha$ can change the vector field such that a left-directed line suddenly changes to a right-directed line.
This is exemplified by the transition from $\alpha=0.4$ to $\alpha=0.5$ on Fig.~\ref{fig:QNMdrift}b where for $\alpha=0.4$, the line emerging from $\omega^{\varepsilon=0}_{n=0}$ ($\omega^{\varepsilon=0}_{n=1}$) is directed left (right), but switches direction when increasing to $\alpha=0.5$.
Although the trajectories are discontinuous when varying $\alpha$, and similar behaviour would occur when varying $x_b$, what matters for us is that for $\alpha$ and $x_b$ fixed, varying $\varepsilon$ defines a continuous map from the open system resonances to the QBS and the new overtones of the semi-open system.
Hence, $\varepsilon$ is a dial which allows us to smoothly switch on this additional structure.

\begin{figure*}
    \centering
    \includegraphics[width=\linewidth
]{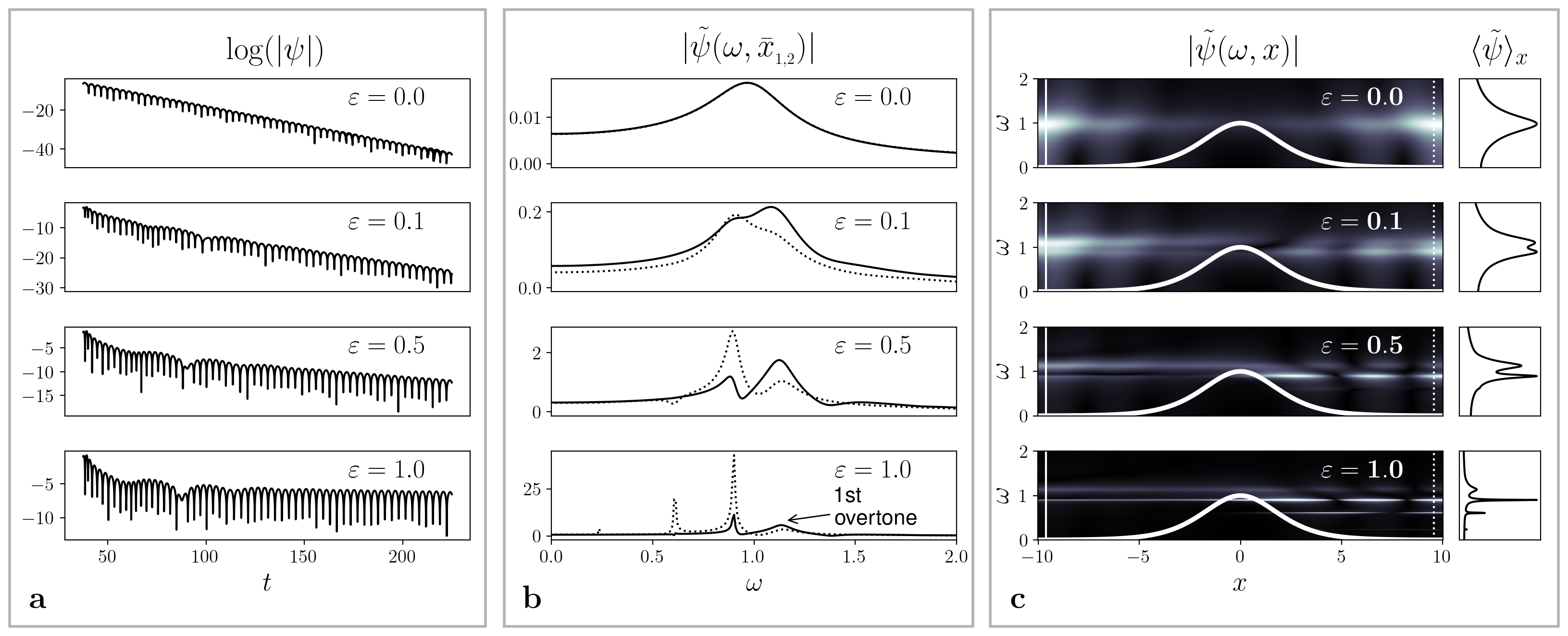}
    \caption{\textbf{a.} Ringdown signal in time and frequency domain for the same $x_b$ and $\alpha$ in Fig.~\ref{fig:QNMdrift}a. The logarithm of the absolute value of the signal close to the reflective wall is shown on the left panels. For the open-system $\varepsilon=0$, we observe the characteristic QNM response whilst for $\varepsilon\neq 0$, the signal is characterised by a fast (slowly) decaying signal at early (late) times. \textbf{b.} We show Fourier transform the of the signal at {$\bar{x}_1$} near the wall (dotted line) and on the open side of the system {at $\bar{x}_2$} (solid). The first overtone (the peak to the right of $\mathrm{Re}[\omega]>1$) becomes more pronounced for larger $\varepsilon$ since the damping is reduced relative to $\varepsilon=0$, and corresponds to the short-lived part of the time domain signal.
    \textbf{c.} We plot the Fourier transform of the signal at each point to illustrate the spatial features for the resonant modes against the PT potential (thick white line). The signal is normalised such that the colour bar goes from 0 (black) to 1 (white). The dotted and the solid lines label $\bar{x}_1$ and $\bar{x}_2$ respectively. The spectrum averaged over the $x$ axis is showed on the right side.}
    \label{fig:sim_pulse}
\end{figure*}

\emph{Numerical simulations.---}A consequence of the reflective boundary is that the new overtones have distinct real frequencies and longer lifetimes, facilitating their detection. We demonstrate this via a numerical simulation.
The wave equation \eqref{eq:WaveEq} is simulated using a method of lines algorithm. Spatial derivatives are implemented using 3-point finite-difference stencils and the evolution is performed using a fourth-order Runge-Kutta algorithm, whilst the permeable wall is implemented by specifying a relation between the $x$ and $t$ derivatives at one of the boundaries (see Appendix B for details). In Fig.~\ref{fig:sim_pulse}, a Gaussian pulse $\exp(i k_0 x-(x-x_0)^2/2\sigma^2)$, with $k_0=2$, centred at  $x_0=-7.5$ (near the open boundary), and  $\sigma^2 =  5$ is evolved toward the barrier.
We illustrate the response signal, in both the time and frequency domains (left and right panels respectively) after the initial pulse has passed for various $\varepsilon$, i.e. as the boundary is made more reflective, focusing on values which make the decay times of $n=0,1$ comparable (see Fig.~\ref{fig:QNMdrift}).
For the open system, we observe the characteristic QNM waveform, corresponding to a singly peaked frequency spectrum. 
However, for semi-open systems, the signal becomes more complicated, characterised by interference between multiple frequencies in the time domain which correspond to more than one peak in Fourier space. 
Notably, the peak corresponding to $\omega^{\varepsilon=0}_{n=0}$ splits in two, the left peak corresponding to the highest frequency QBS and the right peak to the first of the new overtones.
For a partially absorbing wall, we observe that this overtone is actually the dominant signal when we measure near the open boundary.
When the wall is fully reflective, the dominant signal comes from the QBS due to their long lifetimes, although our simulations suggest that the first overtone is still detectable.
In Fig.~\ref{fig:sim_pulse}c, we illustrate the full spatial waveform as a function of frequency, illustrating how the different resonant modes (indicated by bright horizontal lines) are bound by the potential barrier. As expected, modes scattering with the wider part of the barrier experience less leakage. Note that although format in Fig.~\ref{fig:sim_pulse}a will be familiar to gravitational community, the equivalent representations in panels b and c emerge naturally in the data analysis of analogue experiments \cite{Torres2020,aasvancara2024} and carry more information about the spatial profile of the modes.

\begin{figure}
    \centering
    \includegraphics[width=1\linewidth]{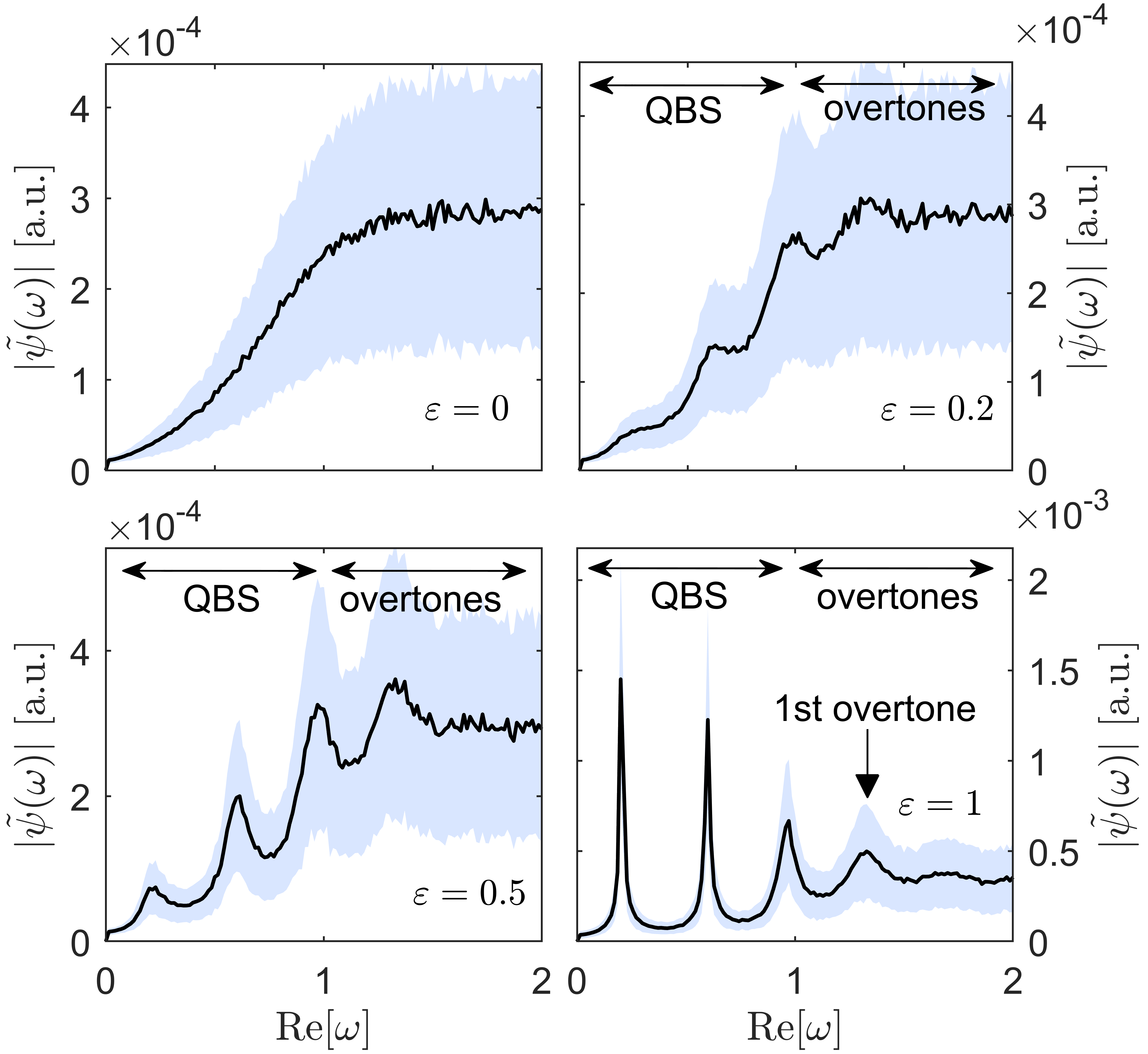}
    \caption{The systems response in the frequency domain when stimulated with mechanical noise. We take the average over 500 runs (black lines) and display the standard deviation as the shaded regions. The semi-open system is characterised by the presence of well defined resonance peaks. In particular, we manage to resolve the first overtone above the noise level. The data is for a potential with $\alpha=1$ and a wall at $x_b=10$.
    }
    \label{fig:noise}
\end{figure}

Although Gaussian pulses are frequently used in numerical studies to stimulate the system \cite{Berti2009}, a more natural perturbing mechanism from the experimental perspective is broadband mechanical noise \cite{Torres2020,svancara2024}.
In Fig.~\ref{fig:noise}, we take $\Psi(x,t=0) = 0$ as the initial condition but generate random noise on the side of the open boundary at each time step for the duration of the simulation. We then Fourier transform the signal after the noise has propagated across the system and take the average on the side of the barrier near the wall.
We collect statistics over 500 iterations and display the mean and the variance.
In the open system, there is no distinct signal; at high frequency we observe the noise level and at lower frequencies a suppression due to the presence of the barrier.
For partial absorption, we observe a series of peaks emerging above the noise level, which become very pronounced for the fully reflective wall.
Notably, the first overtone appears as a distinct signal in the data, suggesting that the combination of a finite size system with mechanical noise may be sufficient ingredients to detect QNM overtones in analogue experiments.
This method of resonant stimulation is akin to that of the driven, damped harmonic oscillator, where the energy input from the driving force is balanced by the energy loss due to dissipation (in our case resulting from the open boundary).

\emph{Discussion.---}Finite-size constraints of analogue experiments might initially appear as an obstacle to simulating black hole ringdown -- a phenomenon typically associated with open (i.e. spatially unbounded) systems. Counter to this expectation, our results demonstrate that the introduction of such confining mechanisms, in fact, becomes a valuable tool for studying the quasinormal modes (QNMs) and the role of the overtones in shaping the ringdown signal. The set-up we have presented (based on a P\"oschl-Teller potential) captures the essential physics of experimentally realisable black hole analogues, where the partially reflective boundary models the experimental confinement and the open boundary represents the analogue black hole horizon. As the reflectivity of the closed boundary is increased from zero, the QNMs of the open system migrate continuously along trajectories in the complex frequency plane toward the real axis. In addition to the familiar quasibound states of semi-open systems, the modified spectrum contains a hierarchy of overtones with distinct oscillation frequencies and substantially longer lifetimes compared to the open system. Qualitatively similar results are expected for any asymptotically flat scattering potential with a single peak, e.g. as encountered around Schwarzschild black holes \cite{vishveshwara1970scattering}.

These two features significantly increase the prospect of detecting QNM overtones in an analogue experiment, as exemplified by our numerical simulations. 
An added consequence is that the frequency response of the semi-open system to a random driving signal is sharply peaked around the QNMs, a feature which is absent in the open system as it relies on constructive interference with reflected waves. A measurement of such a signal would be particularly natural for experiments where noise (e.g. thermal or mechanical) is always present. Indeed, we recently performed experiments on a confined draining vortex in superfluid $^4$He \cite{smaniotto2025}, using the experimental noise to stimulate resonant spectral peaks. This mechanism may also be responsible for the QNM oscillations measured in classical fluid experiments \cite{Torres2022}.

Besides their astrophysical relevance, QNMs are important in a wide variety of subfields, including non-Hermitian systems \cite{leung1994completeness}, strongly coupled field theories \cite{horowitz2000quasinormal,son2007viscosity},
optical cavities \cite{dalton2001theory,kristensen2015normalization}, 
semiclassical gravity \cite{hod1998bohr,maggiore2008physical} and quantum vortex dynamics \cite{patrick2023stability,lan2023splitting}.
By connecting QNMs of the fully open system with those of the semi-open one, we anticipate our results will inspire new crossovers between black hole perturbation theory and the areas mentioned above.

From a gravity perspective, the semi-open problem has been addressed in the context of anti-de Sitter spacetimes, compact objects and neutron stars.
Long-lived modes are known to develop in all these cases \cite{Berti2009,pani4,kokkotas1996pulsatingrelativisticstars}, with the QBS and new overtones in our model mirroring w-modes of neutron stars called trapped and curvature modes. A finite-sized analogue system with variable reflectivity could therefore be used to experimentally investigate the transition between the neutron star regime (full reflection) and the compact object regime (partial reflection), using mechanical noise in the set-up as a probe of the spectrum. Stimulating QNMs with noise may also be relevant in purely astrophysical settings, such as black holes surrounded by matter \cite{barausse2014can,cole2023distinguishing,cannizzaro2024impact} or hypothetical ultra-light fields \cite{hannuksela2019probing,baumann2022sharp}.

Discerning the roles of nonlinearities and overtones in shaping the ringdown waveform is a key topic in gravitational wave research \cite{london2014modeling,mitman2023nonlinearities}, and understanding their interplay is highly desirable. Finite-sized black hole simulators provide a valuable platform for this, allowing experimental control over both nonlinearities (via signal amplitudes) and overtone lifetimes (through boundary reflectivity)--a capability absent in gravitational black holes, where the no-hair theorem fixes overtone lifetimes \cite{Berti2009}. Moreover, such analogue systems enable the study of complex nonlinear interactions relevant to black hole physics, including QCD-axions \cite{yoshino2012bosenova} and photon-plasma effects \cite{cannizzaro2024nonlinear}. While previous studies have explored basic nonlinear terms, e.g., quartic self-couplings \cite{baryakhtar2021black}, gravity simulators offer the potential to probe higher-order interactions experimentally, a direction we will pursue in future work.

We conclude by emphasising that typical methods of stimulating QNMs with wave packets fail in confined systems, since reflections from the boundary obscure the signal, as revealed by earlier experiments \cite{Torres2020}. Furthermore, since real experiments are intrinsically messy and overtones are short-lived, resolving them above the noise can prove difficult. Whilst this would suggest that finite-size effects and experimental noise are obstacles to the experimental study of QNMs, our results demonstrate that these features can actually be used to the experimentalist’s advantage.

\emph{Acknowledgements.---}L.S. and S.P. would like to thank Th\'eo Torres for providing valuable insight on spectral stability based on unpublished work with Sam Dolan.
We are grateful to Patrik \v{S}van\v{c}ara, Pietro Smaniotto and Thomas Sotiriou for helpful discussions and comments on the manuscript. We also thank Paolo Pani, Kostas Kokkotas and Sebastian V\"{o}lkel for feedback. L.S. and S.W. gratefully acknowledge the support of the Leverhulme Research Leadership Award (RL-2019-020). S.P., R.G. and S.W. extend their appreciation to the Science and Technology Facilities Council for their generous support in Quantum Simulators for Fundamental Physics (QSimFP), (ST/T005858/1, ST/Y004450/1, and ST/T006900/1), as part of the UKRI Quantum Technologies for Fundamental Physics program. S.W. also acknowledges the Royal Society University Research Fellowship (UF120112). R.G. acknowledges support from the Perimeter Institute. Research at Perimeter Institute is supported by the Government of Canada through the Department of Innovation, Science and Economic Development Canada and by the Province of Ontario through the Ministry of Research, Innovation and Science.

\bibliography{export}

\section*{Supplemental Material}
\appendix
\numberwithin{equation}{section}
\section{P\"oschl-Teller resonance condition}
\label{app:asymp}

The linear wave equation \eqref{eq:WaveEq} with the P\"oschl-Teller potential reads:
\begin{align}
    \frac{\partial^2\psi}{\partial x^2} - \left(\omega^2-\frac{V_0}{\cosh(\alpha x)^2}\right)\psi = 0\,.
\end{align}
The equation can be reduced to an hypegeometric differential equation \cite{Florentin1966}:
\begin{align}
    (1-y)y\frac{\partial^2\phi}{\partial y^2} + [c-(a+b+1)y]\frac{\partial \phi}{\partial y}-ab\phi = 0\,,
\end{align}
whose solution is \eqref{eq:ptsolution}, by introducing the variable:
\[
\phi = \psi\left((1-y)y\right)^{i\omega/2\alpha}\,.
\]
The parameters $a,b,c$, and the spatial variable $y$ are given in the main text. In order to impose the radiating boundary condition \eqref{eq:FreeBound} at $x \to -\infty$, one sees that for large negative positions $y^{\pm i \omega/2\alpha} \to e^{\pm i \omega x}$. Hence, it is possible to identify the right-moving contribution in \eqref{eq:ptsolution} and then to impose the boundary condition on the left with $B=0$.

To look at the behaviour of the solution as $x\to+ \infty$, i.e. $y\to 1$, one can use the transformation rule of the hypergeometric functions \cite{Florentin1966} to get
\begin{widetext}
    \[
    \psi =  
    Ay^{-i\omega/2\alpha}\left[ 
    (1-y)^{i\omega/2\alpha}\frac{\Gamma(c)\Gamma(a+b-c)}{\Gamma(a)\Gamma(b)}{}_2F_1(c-a,c-b,c^*,1-y) +(1-y)^{-i\omega/2\alpha} \frac{\Gamma(c)\Gamma(c-a-b)}{\Gamma(c-a)\Gamma(c-b)}{}_2F_1(a,b,c,1-y)
    \right]\,,
    \]
\end{widetext}
We can identify the out-going and in-going contributions as $x\to \infty$ by noticing that
\[
(1-y)^{\pm i\omega/2\alpha} \xrightarrow{x\gg 1} e^{\mp i\omega x}\,. 
\]
Therefore, one can define a Reflection Coefficient for the PT potential as the ratio between the out-going and in-going contribution at the right of the potential, resulting in \eqref{eq:ptref}

\section{Numerical simulations}
\label{app:num}

We simulate \eqref{eq:WaveEq} using the method of lines. First we define a spatial grid $x_k$ with $k=1...N$ and \mbox{$h = x_{k+1}-x_k$} and a 3-point centred difference stencil for the second derivative,
\begin{equation}
    \partial_x^2\Psi_k = \frac{\Psi_{k+1}-2\Psi_k+\Psi_{k-1}}{h^2}.
\end{equation}
We then define $\Pi=\partial_t\Psi$ and cast \eqref{eq:WaveEq} in the form,
\begin{equation}
    \partial_t \begin{pmatrix}
        \Psi \\ \Pi
    \end{pmatrix} = \begin{pmatrix}
        0 & 1 \\ \partial_x^2-V & 0
    \end{pmatrix}\begin{pmatrix}
        \Psi \\ \Pi
    \end{pmatrix},
\end{equation}
and use a fourth-order Runge-Kutta algorithm to update the field at the next time step.
This method allows us to implement absorbing boundary conditions in a simple way.
At the edge of the domain where $V\simeq0$ our solution will be either a left-moving or right-moving plane wave, whose derivatives satisfy \mbox{$\partial_x\Psi = \pm\partial_t\Psi$}. 
At the same order of accuracy in $h$, the 3-point centred difference stencil for the first derivative,
\begin{equation}
    \partial_x\Psi_k = \frac{\Psi_{k+1}-\Psi_{k-1}}{2h},
\end{equation}
gives \mbox{$\Psi_0 = \Psi_2 -2h\Pi_1$} at the left boundary and \mbox{$\Psi_{N+1} = \Psi_{N-1} -2h\Pi_N$} at the right boundary.
Hence, we modify our stencil for $\partial_x^2$ on the boundaries,
\begin{equation}
    \partial_x^2\Psi_1 = \frac{2\Psi_2-2\Psi_1}{h^2} - \frac{2\Pi_1}{h}.
\end{equation}
and similarly for the point $k=N$.
Note that without the term proportional to $\Pi_1$, this would just be the second derivative implemented with Neumann boundary conditions, say ${}^\mathrm{N}\partial_x^2$.
Our numerical equation of motion then gets modified to,
\begin{equation} \label{num_eq}
    \partial_t \begin{pmatrix}
        \Psi \\ \Pi
    \end{pmatrix} = \begin{pmatrix}
        \mathbf{0} & \mathbf{I} \\ {}^\mathrm{N}\partial_x^2-V & B
    \end{pmatrix}\begin{pmatrix}
        \Psi \\ \Pi
    \end{pmatrix},
\end{equation}
where $B = -(2/h)\mathrm{diag}(1,0,...0,\mu)$ is a diagonal matrix with non-zero entries in the top-left and bottom-right corner.
The parameter $\mu$ allows us to tune between a fully reflective Neumann boundary condition on the right grid boundary for $\mu=0$ and a completely open boundary for $\mu=1$.
The relation between $\mu$ and the reflectivity $\varepsilon$ used in the main text is,
\begin{equation}
    \mu = \frac{1-\varepsilon}{1+\varepsilon}.
\end{equation}

To generate noise, we add a source term to the right-hand side of Eq.~\eqref{num_eq} in the form of $(\mathbf{0}\ J(x))^\mathrm{T}$. We generate random noise on the open side of the system by setting $J(x)$ everywhere equal to zero except at the point $x_5$ where we set $J(x_5)$ equal to a different random number between $-1$ and $1$ at each time step.

\end{document}